# Electric field manipulation of the Dzyaloshinskii–Moriya Interaction in hybrid multiferroic structures

O.G. Udalov[1,2], R.V. Gorev[1], N.S. Gusev[1,3], A.V. Sadovnikov[4,5], M.V. Sapozhnikov[1,3]

[1]*Institute for Physics of Microstructures RAS, Nizhny Novgorod, GSP-105, Russia*
[2]*California State University Northridge, Northridge, CA, 91330, USA*
[3]*Lobachevsky State University, Nizhny Novgorod 603950, Russia*
[4]*Saratov State University, Saratov 410012, Russia*
[5]*Laboratory of Spin-Orbitronics, Institute of High Technologies and Advanced Materials, Far Eastern Federal University, Vladivostok 690922, Russia*

Hybrid multiferroic films are fabricated by depositing of Pt/Co/Pt multilayers onto [001] and [110] cuts of PMN-PT crystal. The dependence of the interfacial Dzyaloshinskii–Moriya interaction (iDMI) on applied electric field is experimentally investigated in the system by the Brillouin light scattering method. A strong variation (from -0.2 to 0.8 mJ/m$^2$) of the iDMI constant is observed when the electric field is applied. In the case of [001] cut, the observed changes in the iDMI have an isotropic character, while in the case of [110] cut they are anisotropic, which corresponds to the symmetry of the PMN-PT deformations. The change in the iDMI is accompanied by the formation of various unusual domain structures and skyrmion lattices. External control of the DMI with an electric field opens the way to manipulate topological magnetic solitons (such as skyrmions), which are promising object for information processing and storage.

## I. Introduction

The Dzyaloshinskii-Moriya interaction (DMI) attracts significant attention nowadays among groups working in the field of magnetism [1-4]. This is due to the fact that the DMI is responsible for formation of numerous interesting magnetic nanostructures or solitons [5-7]. The most prominent soliton is a magnetic skyrmion – 2D magnetic texture with non-zero topological charge. Skyrmions demonstrate high stability due to topological protection and are proposed to be used in the next generation of magnetic race-track memory [8]. They also demonstrate several interesting physical phenomena, such as topological Hall effect [9,10] and skyrmion Hall effect [11,12]. Recently, 3D solitons such as torons and hopfions were proposed and observed [13,14]. These structures also appear due to the DMI. Interestingly, the DMI appears at an interface between ferromagnetic (FM) and heavy-metal (HM) films [15-17]. This allows to create magnetic layers with predefined and tunable DMI and opens the way to create magnetic solitons with desired properties [18,19]. Such multilayers are in focus of many research groups at the moment [15-21].

Note that the topological solitons are considered as promising candidates for the next generation information storage and processing systems. Control over the DMI interaction opens the way to control these solitons. Therefore, studying the DMI and the ways to modify it has strong application aspect.

While the DMI can be predefined during the fabrication process of multilayer structures, the control of the DMI with some external stimulus is another interesting topic. Such a control allows to manipulate magnetic nanostructures (create, annihilate or move skyrmions, for example). Few approaches were proposed so far to control the DMI in multilayer films. The DMI can be tuned by electric field in the insulator (I)/FM bilayer [22-28]. In this case the electric field produces charge accumulation at the interface changing the DMI value. Another approach is to use mechanical strain to tune the DMI [29-33]. Such a strain induces interatomic distance variation leading to the DMI value changing [29,33].

Interestingly, the charge accumulation mechanism leads to isotropic variation of the DMI. Initially isotropic DMI stays isotropic after the application of an electric field.

In the field of magnetoelectric phenomena, there is a well-developed approach to transform electric field into mechanical strain and then to a magnetic property variation through a coupling of the two effects: electrostriction and magnetostriction [34-38]. Magnetic layered structure can be deposited on top of a piezoelectric (or ferroelectric (FE)) crystal. Applying electric field to the piezoelectric material creates mechanical strain transferred into the magnetic film. This strain induces magnetostriction effect – which is magnetic anisotropy variation under an applied mechanical deformation.

In our previous studies [29, 33] we investigated the DMI transformations under the uniaxial strain caused in the Co/Pt film by the mechanical substrate bending. It was demonstrated that the DMI coefficient becomes anisotropic and even can be made of different sign along different axes in this case. In the present work we study the DMI and magnetic states of the similar Co/Pt films under the stress of the other symmetry, namely uniform and biaxial deformations. For this purpose, we sputter the studied films onto the surface of the FE crystal. We also show that in the hybrid structure FE/buffer layer/FM/HM one can use electric field control the DMI through the mechanical strain. In contrast to the charge accumulation mediated DMI control [22-28] the strain mediated one allows to control not only the DMI value but also the DMI anisotropy. Depending on the electric field direction (with respect to the FE crystal axis) one can induce either isotropic strain or anisotropic strain leading to either isotropic or anisotropic DMI variation. This may eventually allow to manipulate skyrmions or other magnetic textures with an electric field – create and annihilate magnetic solitons or even move them along the film. Strain induced skyrmion motion was recently studied theoretically [39]. Such an approach has some advantages (for example, essential reduction of the skyrmion Hall effect) comparing to the current driven skyrmion motion.

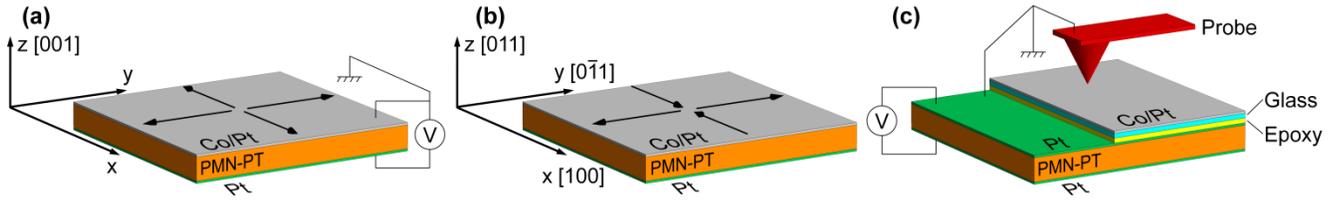

Fig. 1. (Color online) The schematic representation of the samples and voltage application circuit. (a) The [001] CoPt/PMNPT hybrid multiferroic structures. (b) The [110] CoPt/PMNPT hybrid multiferroic structures. Black arrows show the in-plane deformation directions. (c) The sample with a Co/Pt multilayer glued on the top of PMNPT crystal for MFM investigations.

The paper organized as follows. In Sec. II the samples fabrication is described. In Sec. III the main method for the electric field induced DMI measurement – the Brillouin light scattering (BLS) is described. In Sec. IV the results on the electric field induced DMI variation are presented and finally in Sec. V we use a magnetic force microscopy (MFM) to show how the electric field variation of the DMI affects a magnetic domain structure of the system.

## II. Samples fabrication

The hybrid multiferroic structures is fabricated by a deposition of thin Pt/Co/Pt films on the top of a PMN-PT crystal by a dc magnetron sputtering in Ar atmosphere. The thickness is 1.9 nm for the Pt layers and 1.3 nm for Co. The parameters of the films are chosen according to [29] to achieve strong strain dependent DMI variation. The growth rate is 0.25 nm/s for Pt and 0.125 nm/s for Co. The accuracy of the deposition of the layers in thickness is ~10%. The commercially available (Atom Optics Co., Ltd., China) PMN-PT crystal with the composition of 68%Pb($Mg_{1/3}Nb_{2/3}$)$O_3$ - 32%$PbTiO_3$ are used as substrate for the CoPt films. The crystal has the form of a polished plane-parallel plate of 0.5 mm in thickness. The lateral size of plate is about 5 mm. A Pt(10nm)/Ta(90nm) film is deposited onto the bottom plane of the PMN–PT crystal. It serves as the lower electrode. Two different cuts of PMN-PT ([001] and [011]) are used to obtain isotropic and anisotropic in-plane deformations of the ferroelectric by applying electrical field (Fig. 1a,b).

The resulting magnetic film has a perpendicular magnetic anisotropy. A number of samples were examined using magneto-optical Kerr effect (MOKE) measurements to select those in which perpendicular anisotropy exists but is low leading to weak pinning of the domain walls (Fig. 1b). In such samples, the residual magnetization is smaller, they are more sensitive to iDMI variations [29,33]. As the result the Co thickness in the selected samples is chosen so, that the residual magnetization in the films is approximately 15% of the saturation. The samples demonstrate low coercivity which depends on the applied electrical field (Fig. 2b).

These structures are used to measure the dependence of the DMI on applied voltage. Unfortunately, even the polished PMN-PT crystals at our disposal have a rather rough surface. Atomic force microscopy (AFM) shows the anisotropic nature of the surface morphology (Fig. 2c). As measurements show, the surface unevenness is gentle, the maximum slope is no more than 10%. This cannot lead to film ruptures; clustering is not observed even in Co layers 0.5 nm thick [40]. Nevertheless, magnetostatic interactions impose an anisotropic stripe domain structure on the magnetic film in this case. This does not interfere with BLS measurements, since they are carried out in a strong saturating field, but the pinning of the domains is too strong, which makes it impossible to see any change in the magnetic structure upon application of voltage. Therefore, an alternative series of samples are fabricated for magnetic force microscopy (MFM) measurements. To obtain a smooth surface, a magnetic film is sputtered onto a glass substrate (0.18 mm thick), which is then glued to the PMN-PT crystals using an epoxy. Prior to the gluing, the crystal is covered with PtTa films on both top and bottom surfaces. These layers act as electrodes for voltage application. In this case, the magnetic film and MFM tip are always grounded. This structure is schematically shown in Fig. 1(c).

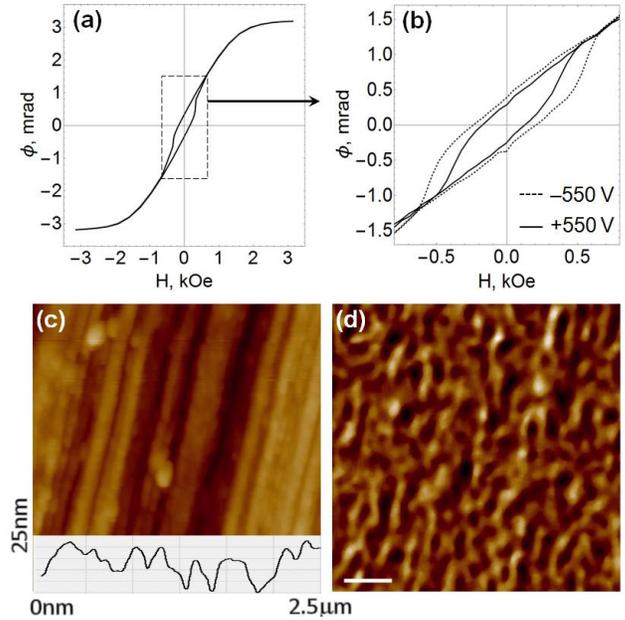

Fig. 2 (Color online) (a) MOKE of our typical CoPt film in the perpendicular magnetic field. (b) The dependence of the MOKE hysteresis loop on the applied voltage. (c) Typical AFM image of the CoPt/PMN-PT structure the corresponding cross-section is shown. (d) The MFM image of the same sample.

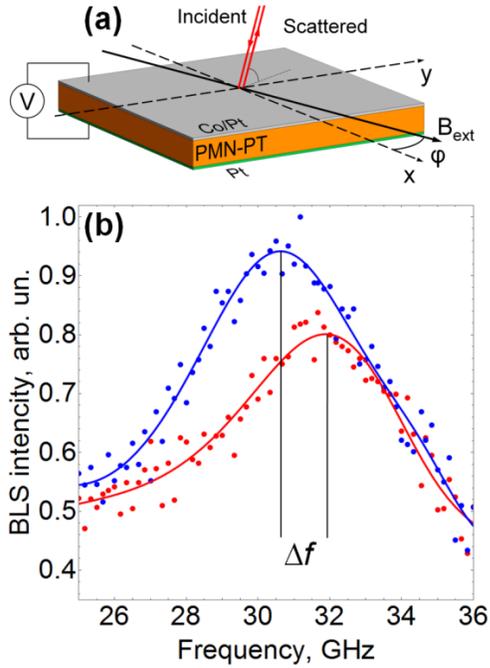

Fig. 3.(Color online) (a) BLS measurement geometry. (b)Typical BLS spectrumof the sample Glass/Ta/Pt/Co/Pt without a strain at H = 1 T (blue dots)and H = −1 T (red dots). Solid lines are Lorentzian fits. Δf is the frequency shift between the Stokes and anti-Stokes peaks.

### III. BLS measurements procedure

The DMI in the samples is studied by the Brillouin light scattering (BLS) in the Damon-Eshbach geometry [41] (see Fig. 3a) which corresponds to the spin waves propagating at the 90 deg angle to the equilibrium magnetization direction. A magnetic field is applied in the sample plane along three possible different directions (x, y and diagonal). The BLS intensity reaches maximum when the incident wave (generated by the laser light) is polarized perpendicular to the magnetization and the sample is saturated in the film plane [42]. We observe the BLS signal at magnetic fields$H$>8 kOe meaning that the sample is saturated in-plane at these fields. This is in agreement with the perpendicular direction of the magnetic anisotropy. The DMI constant along each direction is measured depending on an applied electric field. BLS spectra from thermally excited magnons (spin waves (SW)) were measured in the backscattering configuration with a 300-mW single mode laser operating at a wavelength λ = 532 nm with linewidth of 15 MHz. The light was focused on the surface of the CoPt film using the lenses with a numerical aperture of 0.29 and 0.35, resulting in a laser spot diameter of 25 μm and 50 μm, correspondingly. The sample was mounted on a mechanical goniometer in order to choose a specified angle of incidence of the light (θ). Backscattered light from SWs has a polarization rotated by 90 degrees with respect to the laser light polarization. The scattered light passes through the analyzer in order to suppress the signal from elastically scattered light and to minimize the contribution of the surface phonon. Since the in-plane momentum should be conserved the variation of θ leads to select the magnitude of the in-plane component of the SW wave vector. The electric field is applied perpendicular to the sample plane (see Fig. 1a). The bottom electrode is a nonmagnetic film deposited on the PMN-PT crystal surface. The top electrode is the magnetic multilayer itself. Voltage is applied to the bottom electrode while the top magnetic film (the sample) is grounded. Application of the electric field induces different kind of deformation in the plane of the FE crystal (xy-plane). The strain is isotropic for the [001] cut and anisotropic for [011]. Importantly, the in-plane strain is transferred to the magnetic film when one applies the electric field.

In the BLS experiment we sweep the voltage between -500 V to + 500 V, which gives the field variation of ± 1 MV/m. We sweep the field few times and then average the data over these sweeps. Two different sizes of the laser spot are used in the BLS measurements: 25um and 50 um. No essential difference is observed in these two experiments. The size of the laser spot is much bigger than the magnetic domains and bigger than the FE domains. The BLS spectrums are obtained from few different spots on the sample. No essential difference is observed.

The BLS signal contains the information about the frequency of the scattered light, which was analyzed using a (3+3)-tandem Fabry-Pérot interferometer. Typical BLS spectrum is presented in Fig. 3 (b). Solid lines show the Lorentzian fit demonstrating the shift of the Stokes and anti-Stokes peaks denoted as Δf. Following the standard approach, we estimate the DMI constant as [41 and 43]

$$D_i = 2M_s \Delta f/(\pi \gamma k_i), \qquad (1)$$

where $M_s$ is the saturation magnetization, $\gamma$ = 176 GHz/T is the gyromagnetic ratio, $\Delta f$ is the difference between the Stokes and anti-Stokes frequencies, and $k_i$ is the momentum along the i-direction (in our case i = x or y or the direction along diagonal axis). The value of $M_s$ used in our estimations is $1.1 \times 10^6$ A/m which is typical for Co/Pt films [44,45].

### IV. BLS measurements results

The results for the Co/Pt multilayers fabricated on top of the isotropic [001] PMN-PT crystal are shown in Fig. 4. One can see essential variation of the DMI constant with electric field. It changes from 0.6 to -0.2 mJ/m$^2$.Other researches saw the FE hysteresis in this field range [46]. We observe very narrow hysteresis behavior, as it is seen in Fig. 4a.
Interestingly, that the DMI constant changes sign when positive voltage is applied. Fig. 4a shows the DMI constant along the x-direction and Fig. 4b shows the DMI constant along the y-direction. One can see that they approximately coincide (within the error bar). So, the DMI constant is isotropic. This is in agreement with the strain symmetry. As we mentioned above the electric field induces isotropic in-plane strain in PMN-PT crystal with [001] cut leading to isotropic in-plane strain in the magnetic film.

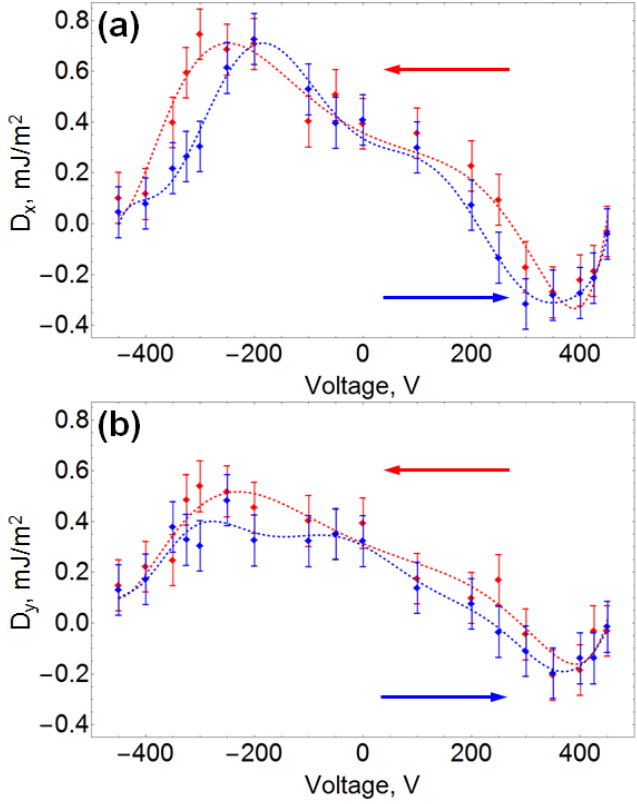

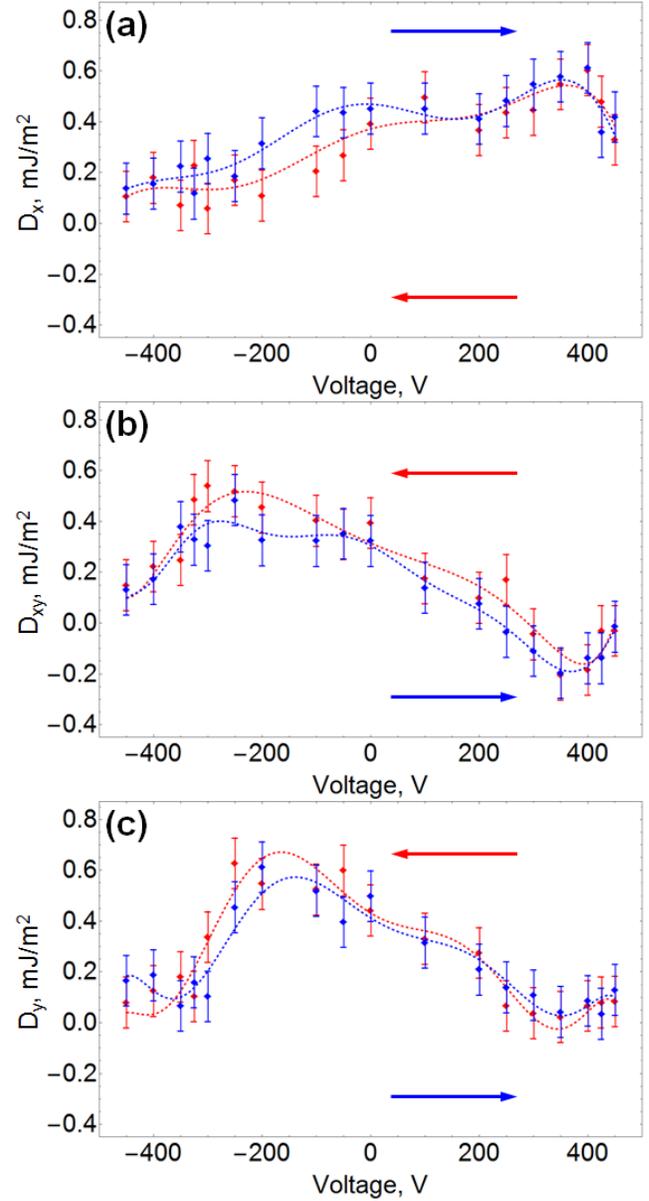

Fig. 4.(Color online)The dependence of the DMI interaction constant on the applied electric field for the Co/Pt film grown on [001] cut of the PMN-PT crystal. (a) $D_x$ measured with the magnetic field applied along x-direction. (b) $D_y$ measured with the magnetic field applied along y-direction. Arrows and corresponding colors denote the direction of the voltage sweep. Dashed lines are the guides for eyes.

    The DMI constant is a non-monotonic and asymmetric function of the strain. This is in agreement with the previous experiments on the strain dependent DMI [29,33]. Recent theoretical consideration of the DMI variation induced by mechanical deformations also showed that the DMI can be the non-monotonic function of the strain [47]. This is related to the fact the DMI origin is similar to the RKKY interaction [17] and therefore it has strong and alternating dependence on the inter atomic distance.

    Fig. 5 shows the results for the Co/Pt film grown on the [011] cut of the PMN-PT crystal. One can see that the DMI constant changes in this case as well. The variation is strong (from 0 to 0.6 mJ/m$^2$). The behavior is non-monotonic as well. Applying electric field to such a crystal induces anisotropic in-plane strain (contraction along one axis and expansion along another). Previously, it was demonstrated that such a strain may induce anisotropic DMI with different DMI

Fig. 5. (Color online) The DMI strength as a function of the applied voltage for the Co/Pt film grown on the [011] cut of the PMN-PN crystal. (a) $D_x$ measured with the magnetic field applied along x-direction. (b) $D_{xy}$ measured with the magnetic field applied along diagonal direction (45 degree with respect to x-axis). (c) $D_y$ measured with the magnetic field applied along y-direction. Arrows and corresponding colors denote the direction of the voltage sweep. Dashed lines are the guide for eyes.

constants along different axes [29]. Experimental data shown in Fig. 5 confirms this effect. In the absence of the electric field the system is isotropic and the DMI constant is the same along x- and y-directions ($D_x = D_y \sim 0.4$ mJ/m$^2$). When the electric field is applied, the DMI constant along x-direction $D_x$ becomes different than the y-component $D_y$. Note that the dependencies of the $D_x$ variation and $D_y$ variation are opposite. At low voltages ($|V| < 100$ V), $dD_x/dV < 0$ while $dD_y/dV > 0$. This is an expected behavior since the strain behaves in the same way: $d\varepsilon_{xx}/dV > 0$ and $d\varepsilon_{yy}/dV < 0$. One can see that the anisotropy of the DMI is quite high. At

$V = 400$ V, $D_x \sim 0.5$ and $D_y \sim 0$ and the opposite situation appears at negative voltage.

Figure 5b shows the DMI constant along the diagonal direction. One can expect that the DMI along the diagonal direction is an average of $D_x$ and $D_y$. In this case the effect of the anisotropic strain should be reduced. However, in fact the DMI in the anisotropic system has an additional component [48]. Below is the expression for the DMI energy in the coordinate system rotated by angle β with respect to a principal coordinate system.

$$W_{DMI} = D_{x'}\left(m_{x'}\frac{\partial m_z}{\partial x'} - m_z\frac{\partial m_x}{\partial x'}\right) +$$
$$+D_{y'}\left(m_{y'}\frac{\partial m_z}{\partial y'} - m_z\frac{\partial m_y}{\partial y'}\right) +$$
$$+D_\perp\left(m_{x'}\frac{\partial m_z}{\partial y'} + m_{y'}\frac{\partial m_z}{\partial x'} - m_z\frac{\partial m_{y'}}{\partial x'} - m_z\frac{\partial m_{x'}}{\partial y'}\right) \quad (2)$$

where

$$D_{x'} = D_x \cos^2(\beta) + D_y \sin^2(\beta)$$
$$D_{y'} = D_x \sin^2(\beta) + D_y \cos^2(\beta) \quad (3)$$
$$D_\perp = (D_x - D_y) \cos(\beta) \sin(\beta)$$

In our case the principal coordinate system is the one oriented along the main strain axes. To get the DMI energy in the coordinate system with the x'-axis oriented by 45 degree with respect to our principal axes (along x=y direction) we use $\beta = \pi/4$ giving

$$W_{DMI} = \frac{D_x+D_y}{2}\left(m_{x'}\frac{\partial m_z}{\partial x'} - m_z\frac{\partial m_x}{\partial x'} + m_{y'}\frac{\partial m_z}{\partial y'} - m_z\frac{\partial m_y}{\partial y'}\right) +$$
$$+\frac{D_x-D_y}{2}\left(m_{x'}\frac{\partial m_z}{\partial y'} + m_{y'}\frac{\partial m_z}{\partial x'} - m_z\frac{\partial m_{y'}}{\partial x'} - m_z\frac{\partial m_{x'}}{\partial y'}\right) \quad (4)$$

One can see that beside the "traditional" term (first term in Eq. (4)), there is an additional term depending on the difference $(D_x - D_y)$. We think namely this term contributes to the BLS spectrum shift and therefore there is an essential dependence of the DMI strength on the voltage along the diagonal direction. The variation magnitude is similar to $D_x$ and $D_y$.

Comparing Figs. 5a and 5c, one can observe that abrupt decrease of the $D_x$ happens at +400 V, while the same abrupt decrease of $D_y$ happens at -300 V. One can also see that the overall slope of $D_x$ is bigger than that of $D_y$. This suggests that the same electric field induces stronger deformations along the x-axis than along the y-axis, leading to stronger variation of the DMI constant $D_x$. This is in agreement with previous observation of the strain behaviour of the [011] cut PMN-PT crystals, where the strain is higher along one axis than along another one. It is due to additional deformations of the crystal along the z axis when stress is applied. According to $\varepsilon_{xx}(V) + \varepsilon_{yy}(V) + \varepsilon_{zz}(V) = 0$, different magnitude of strain along x- and y-direction may also produces some DMI variation along the diagonal direction through $(D_x + D_y)$ contribution. Note that in the isotropic sample ([001] cut) this effect is absent and the DMI behaviour is the same for both x- and y-axes.

Although we obtained our data as a dependence of DMI on the applied voltage, we can estimate the corresponding strain values based on the literature data. Note, that measured values of the electrostriction coefficient of the PMN-PT vary in different works [49]. We chose some average values for our estimate $d_{33} = 1800$ pC/N and $d_{31} = -500$ pC/N. As we mentioned before we use two different crystal cuts ([001] and [110]). For the [001] orientation, the crystal axes coincide with the sample axes. In this case application of the field along the z-axis create strain in the z-direction $\sigma_{zz} = d_{33}E$ and the isotropic strain in the surface plane, $\sigma_{xx} = \sigma_{yy} = -\nu d_{33}E = -0.6 \cdot 10^{-9}E$, where $\nu \approx 0.33$ is the Poisson ratio of the PMN-PT crystal and E is the electric field applied to PMN-PT crystal (E = V/d, d is the PMN-PT thickness). For the applied voltage of 100V we get approximately 0.012% strain. Note that for higher voltages some non-linearities and even polarization switching may occur.

For the case of the [110] cut the crystal axes do not coincide with the sample axes. We will mark the crystal axes with a prime symbol. Applying electric field along z-axis ([110] axis of the crystal), we get anisotropic strain in the sample plane. We have $\sigma_{yy} = (d_{31} + d_{33})E/\sqrt{2}$, $\sigma_{xx} = d_{31}E\sqrt{2}$. For 100V voltage we have $\sigma_{xx} \approx -0.01\%$ and $\sigma_{yy} \approx 0.02\%$.

## V. Magnetic-force microscopy measurements

Beside the BLS measurements we perform magnetic force microscopy (MFM) studies of our sample. This allows us to see if the DMI variation in the films leads to any changes in the film magnetic domain structure when we apply the electric field.

We used the MFM to visualize magnetic states, existing at zero magnetic field. A phase shift of the probe oscillations is registered as an MFM signal. The measurements are performed using both the two-pass tapping mode and the non-contact constant-height mode. Prior to the measurements, the MFM probes were magnetized along the tip axis in a 1-T external magnetic field. The magnetic film and the tip are grounded to avoid electrostatic effects [50]. The scan size is 10×10 μm² (512 × 512 pixels). It allows to obtain appropriate power Fourier spectra. These spectra are used for the analysis of the domain structure symmetry and for estimations of the domain sizes. Prior to each measurement, the sample is demagnetized in a decaying alternating perpendicular external field. This procedure guarantees that the area of the domains with opposite magnetization would be equal.

As it is mentioned above, the samples with direct deposition of a magnetic film on a PMN-PT crystal have an uneven surface, which pins the domain structure. This does not interfere with the BLS measurements, since they are carried out in a saturating magnetic field. However, no changes in the domain structure are observed in such samples upon application of an electric field. In this regard, the MFM measurements are made on an alternative series of samples in which the magnetic film is deposited on a smooth glass substrate and then glued to the PMN-PT crystal (Fig 1c). The glass substrate has a thickness of 200 um (thinner than the FE crystal). Presence of the glass substrate and the glue layer could reduce the strain transferred from the FE crystal to the FM film. So, we expect weaker effect of the electric field on the DMI constant.

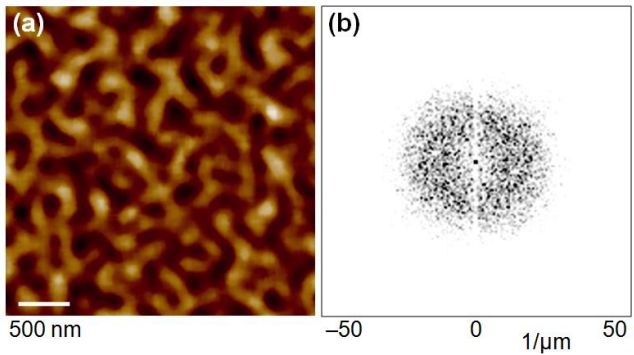

Fig. 6 (Color online) (a) Typical MFM image of the isotropic labyrinth domain structure in the CoPt/glass/[001]PMN-PT structure. (b) Spatial Fourier spectrum of the domain structure. White color corresponds to the Fourier harmonic with zero amplitude. Black color corresponds to the Fourier harmonics with high amplitude. The vertical white line in the center of the diagrams is an artifact of image processing.

The MFM is used to study how the domain structure in our samples changes when an electric field is applied. The measurements show that the restructuring of the domain structure occurs on some sample and does not occur in others. We attribute this to two possible factors. The first one is that the MFM measurements are carried out rather locally (scan size is 10 μm) and, therefore, they are affected by the inhomogeneity of the material parameters of the samples. The second one is that the adhesive epoxy layer may have different thicknesses in different samples and may transfer the strain from the PMN-PT crystal to the magnetic film differently. Below we demonstrate some observed examples of the domain structure variation under the action of the external electric field.

In the CoPt/glass/[001]PMN-PT samples, the in-plane strain produced by the electric field is isotropic. Therefore, the observed labyrinth domain structure (see Fig. 6a) remains isotropic at every voltage. This is confirmed by the cylindrically symmetric (ring type) spatial Fourier spectrum (Fig. 6b). While the application of the electric field does not change the isotropy of the domains, it changes their size. The average domain width is calculated based on the average radial profile of the spatial Fourier spectrum. When the applied voltage changes from $V = -300$ V to $V = +300$ V a decrease in the average size of domains is observed. For one of the samples, the domains size decreases from 175 nm to 145 nm. In another sample, the domains dimensions change from 140 to 120 nm. The change in the size of the domains is due to the change in the energy of the domain walls due to the DMI variations.

CoPt/glass/[011]PMN-PT hybrid structure experiences anisotropic lateral deformations when the voltage is applied. In this case, the changes of the magnetic structure are more prominent. Fig. 8 demonstrates that the isotropic labyrinth domains (observed at $V = 0$ V) transform into an array of magnetic bubbles when the voltage of -300 V is applied. In this case, the lateral Fourier spectrum remains isotropic, but the average domain sizes decrease from 140 to 110 nm. We speculate that due to the strong DMI in the system, the observed magnetic bubbles are actually magnetic skyrmions. The surface profiles obtained in the AFM (Fig. 7a,d) testify that the MFM measurements for different voltages are performed in the same area.

Another interesting example of the domain structure modification is shown in Fig. 8. The MFM images in Fig.8b and 8e corresponds to the same spot at the sample (compare the AFM data of Figs. 8a and 8d). However, they are obtained at different voltages. One can see that isotropic domain structure (Fig. 8e) transforms into a zig-zag domains (Fig. 8b). The Fourier spectrums shown below confirm this transformation. The spectrum in Fig. 8f represents a ring with very weak angular variation of the amplitude meaning that there are no preferable directions for the domains. In contrast, the spectrum in Fig. 8c has 4 spots corresponding to the preferable directions of the domains. Previously, similar zig-zag domains have been observed in Co/Pt magnetic films subjected to a uniaxial lateral mechanical compression [33]. Such a zig-zag domain structure appears due to the DMI anisotropy [48]. Note, that a zig-zag structure cannot be explained by the surface magnetic anisotropy. The magnetic anisotropy may create a single preferred direction but not two preferred directions.

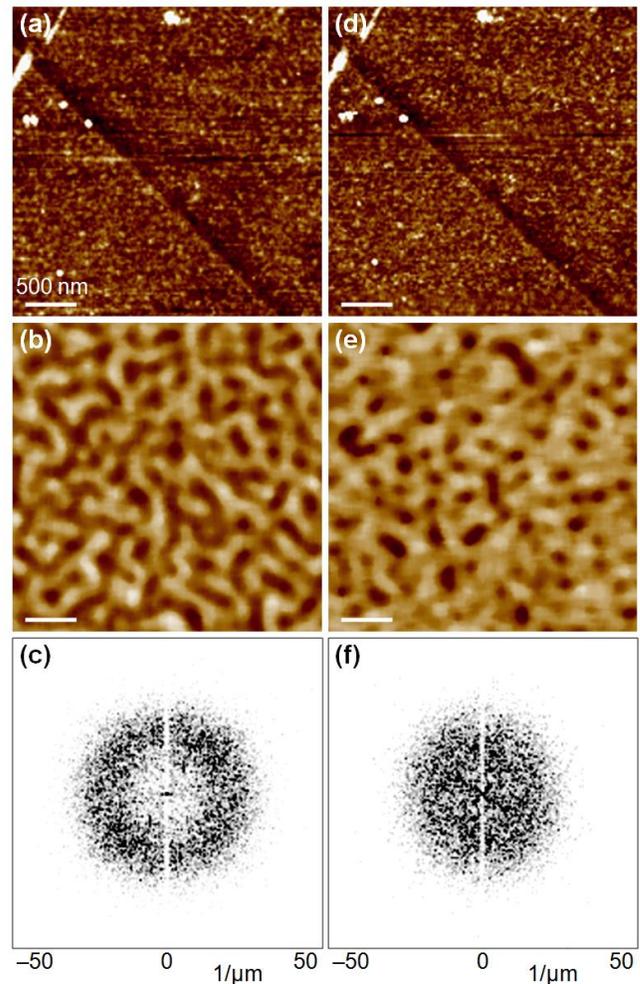

Fig. 7 (Color online) Magnetic configuration observed in the CoPt/glass/[011]PMN-PT structures. (a) AFM image, (b) MFM image and (c) spatial Fourier transforms of the magnetic structure for isotropic labyrinth domains observed at U=0V. (d-f) is the same for magnetic skyrmions, observed at U=-300V.

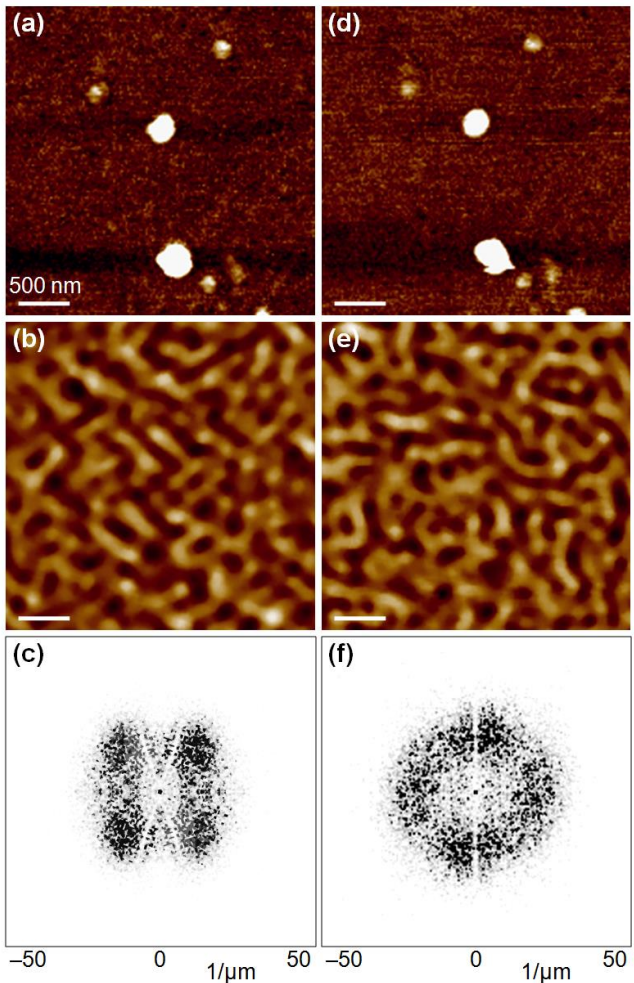

Fig. 8 (Color online) Magnetic configuration observed in the CoPt/glass/[011]PMN-PT structures at two different voltages (a) AFM image, (b) MFM image and (c) spatial Fourier transforms of zig-zag domain structure. (d-f) is the same for isotropic labyrinth domains.

## IV. Conclusion

In the present work we demonstrated that the DMI interaction in a hybrid system Co/Pt/FE crystal can be varied with an electric field applied to the FE crystal perpendicular to its surface. The effect occurs due to appearance of electric field induced mechanical strain transferred from the FE crystal to the FM film through an interface. The variation of the DMI constant is about ± 0.4 mJ/m$^2$ when one applies the electric field of ± 1 MV/m. Such changes are more than 100% variation of the DMI constant, and can lead to a change in the sign of DMI at high enough voltages. Interestingly, depending on the crystal cut the electric field may induce either isotropic or anisotropic DMI variation. For the [011] cut of the PMN-PT crystal the DMI along x direction grows with the field and the DMI along the y-direction decays. The induced anisotropy can be very high with $D_x=0$ and $D_y\sim0.5$ mJ/m$^2$.

## Acknowledgements

The research was carried out within the state assignment of Ministry of Science and Higher Education of the Russian Federation (theme No. FFUF-2021-0021).A.V.S. thanks the Russian Ministry of Science and Higher Education for state support of scientific research conducted under the supervision of leading scientists in Russian institutions of higher education, scientific foundations, and state research centers (project no. 075-15-2021-607).The facilities of Center "Physics and technology of micro- and nanostructures" at IPM RAS is used for the analysis of the samples.

## References

[1] M. Kuepferling, A. Casiraghi, G. Soares, G. Durin, F. Garcia-Sanchez, L. Chen, C. H. Back, C. H. Marrows, S. Tacchi, and G. Carlotti, Measuring interfacial Dzyaloshinskii-Moriya interaction in ultrathin magnetic films, Rev. Mod. Phys. **95**, 015003 (2023).
[2] H. Yang, J. Liang, and Q. Cui, First-principles calculations for Dzyaloshinskii–Moriya interaction, Nat. Rev. Phys.**5**, 43 (2023).
[3] D. Yu, Y. Ga, J. Liang, C.Jia, and H. Yang, Voltage-Controlled Dzyaloshinskii-Moriya Interaction Torque Switching of Perpendicular Magnetization, Phys. Rev. Lett. **130**, 056701 (2023).
[4] Yu-Hao Huang, Chao-Chung Huang, Wei-Bang Liao, Tian-Yue Chen, and Chi-Feng Pai, 4.           Growth-Dependent Interlayer Chiral Exchange and Field-Free Switching, Phys. Rev. Applied 18, 034046 (2022)
[5] A. N. Bogdanovand C. Panagopoulos, Physical foundations and basic properties of magnetic skyrmions, Nat. Rev. Phys.**2**, 492 (2020).
[6] X. Zhang, J. Xia, L. Shen, M. Ezawa, O. A. Tretiakov, G. Zhao, X. Liu, and Y. Zhou, Static and dynamic properties of bimerons in a frustrated ferromagnetic monolayer, Phys. Rev. B **101**, 144435 (2020).
[7] A.Fert, N.Reyren, and V.Cros, Magnetic skyrmions: advances in physics and potential applications, Nat. Rev. Mater.**2**, 17031 (2017).
[8] R. Tomasello, E. Martinez, R. Zivieri, L. Torres, M. Carpentieri, and G. Finocchio, A strategy for the design of skyrmion race track memories, Sci. Rep.**4**, 6784, (2014).
[9] Y. Onose, N. Takeshita, C. Terakura, H. Takagi, and Y. Tokura, Phys. Rev. B **72**, 224431 (2005).
[10] M.V. Sapozhnikov, N.S. Gusev, S.A. Gusev, D.A. Tatarskiy, Yu.V. Petrov, A.G. Temiryazev, and A. A. Fraerman, Direct observation of topological Hall effect in Co/Pt nanostructured films, Phys. Rev. B **103**, 054429 (2021).
[11] W. Jiang, X. Zhang, G. Yu, W. Zhang, X. Wang, M. Benjamin Jungfleisch, J. E. Pearson, X. Cheng, O. Heinonen, K. L. Wang, Y. Zhou, A. Hoffmann, and S. G. E. teVelthuis, Direct observation of the skyrmion Hall effect, Nat. Phys.**13**, 162 (2017).

[12] J. Zang, M. Mostovoy, J. H. Han, and N. Nagaosa, Dynamics of Skyrmion Crystals in Metallic Thin Films, Phys. Rev. Lett. **107**, 136804 (2011).
[13] P. Sutcliffe, Hopfions in chiral magnets., J. Phys. A: Math. Theor. **51**, 375401 (2018).
[14] M. Grelier, F. Godel, A. Vecchiola, S. Collin, K. Bouzehouane, A. Fert, V. Cros, and N. Reyren, Three-dimensional skyrmionic cocoons in magnetic multilayers, Nat.Commun.**13**, 6843 (2022).
[15] H. Yang, A. Thiaville, S. Rohart, A. Fert, and M. Chshiev, Anatomy of Dzyaloshinskii-Moriya Interaction at Co/Pt Interfaces, Phys. Rev. Lett. **115**, 267210 (2015).
[16] Zhiyuan Zhao, Dan Su, Tao Lin, Zhicheng Xie, Duo Zhao, Jianhua Zhao, Na Lei, and Dahai Wei, Separation of Rare-Earth and Transition-Metal Contributions to the Interfacial Dzyaloshinskii-Moriya Interaction in Ferrimagnetic CoGd Alloys, Phys. Rev. Applied 19, 044037 (2023).
[17] A. Fert and P. M. Levy, Role of Anisotropic Exchange Interactions in Determining the Properties of Spin-Glasses, Phys. Rev. Lett. **44**, 1538 (1980).
[18] A. Cao, X. Zhang, B. Koopmans, S. Peng, Y. Zhang, Z. Wang, S. Yan, H. Yang, and W. Zhao, Tuning the Dzyaloshinskii–Moriya interaction in Pt/Co/MgO heterostructures through the MgO thickness, Nanoscale **10**, 12062-12067 (2018).
[19] H. T. Nembach, E. Jué, K. Poetzger, J. Fassbender, T. J. Silva, and J. M. Shaw, Tuning of the Dzyaloshinskii–Moriya interaction by He+ ion irradiation, J. Appl. Phys.**131**, 143901 (2022).
[20] M. V. Sapozhnikov, Y. V. Petrov, N. S. Gusev, A. G. Temiryazev, O. L. Ermolaeva, V. L. Mironov, and O. G. Udalov, Artificial Dense Lattices of Magnetic Skyrmions, Materials**13(1)**, 99 (2020).
[21] A. W. J. Wells, P. M. Shepley, C. H. Marrows, and T. A. Moore, Effect of interfacial intermixing on the Dzyaloshinskii-Moriya interaction in Pt/Co/Pt, Phys. Rev. B **95**, 054428 (2017).
[22] B. H. Zhang, Y. S. Hou, Z. Wang, and R. Q. Wu, Tuning Dzyaloshinskii-Moriya interactions in magnetic bilayers with a ferroelectric substrate, Phys. Rev. B **103**, 054417 (2021).
[23] W. Zhang, H. Zhong, R. Zang, Y. Zhang, S. Yu, G. Han, G. L. Liu, S. S. Yan, S. Kang, and L. M. Mei, Electrical field enhanced interfacial Dzyaloshinskii-Moriya interaction in MgO/Fe/Pt system, Appl. Phys. Lett. **113**, 122406 (2018).
[24] C. Huang, L. Z. Jiang, Y. Zhu, Y. F. Pan, J. Y. Fan, C. L. Ma, J. Hu, and D. N. Shi, Tuning Dzyaloshinskii–Moriya interaction via an electric field at the Co/h-BN interface, Phys. Chem. Chem. Phys.**23**, 22246-22250 (2021).
[25] L. Herrera Diez, Y. T. Liu, D. A. Gilbert, M. Belmeguenai, J. Vogel, S. Pizzini, E. Martinez, A. Lamperti, J. B. Mohammed, A. Laborieux, Y. Roussigné, A. J. Rutter, E. Arenholtz, P. Quarterman, B. Maranville, S. Ono, M. Salah El Hadri, R. Tolley, E. E. Fullerton, L. Sanchez-Tejerina, A. Stashkevich, S.M. Chérif, A.D. Kent, D. Querlioz, J. Langer, B. Ocker, and D. Ravelosona. Nonvolatile Ionic Modification of the Dzyaloshinskii-Moriya Interaction, Phys. Rev. Appl.**12**, 034005 (2019).
[26] M. Schott, L. Ranno, H. Béa, C. Baraduc, S. Auffret, A. Bernand-Mantel, Electric field control of interfacial Dzyaloshinskii-Moriya interaction in Pt/Co/AlO$_x$ thin films, J. Magn. Magn. Mater. **520**, 167122 (2021).
[27] C. Ederer, and C. J Fennie, Electric-field switchable magnetization via the Dzyaloshinskii–Moriya interaction: FeTiO$_3$ versus BiFeO$_3$, J. Phys. Condens. Matter.**20**, 434219 (2008).
[28] T. Srivastava, M. Schott, R. Juge, V. Křižáková, M. Belmeguenai, Y. Roussigné, A. Bernand-Mantel, L. Ranno, S. Pizzini, S.-M. Chérif, A. Stashkevich, S. Auffret, O. Boulle, G. Gaudin, M. Chshiev, C. Baraduc, and H. Béa, Large-Voltage Tuning of Dzyaloshinskii–Moriya Interactions: A Route toward Dynamic Control of Skyrmion Chirality, Nano Lett.**18**, 8, 4871 (2018).
[29] N. S. Gusev, A. V. Sadovnikov, S. A. Nikitov, M. V. Sapozhnikov, and O. G. Udalov, Manipulation of the Dzyaloshinskii–Moriya Interaction in Co/Pt Multilayers with Strain, Phys. Rev. Lett**124**, 157202 (2020).
[30] Z. Xu, Q. Liu, Y. Ji, X. Li, J. Li, J. Wang, and L. Chen, Strain-Tunable Interfacial Dzyaloshinskii–Moriya Interaction and Spin-Hall Topological Hall Effect in Pt/Tm$_3$Fe$_5$O$_{12}$Heterostructures, ACS Appl. Mater. Interfaces**14**, 14, 16791 (2022).
[31] A. Ebrahimian, A. Dyrdał, and A. Qaiumzadeh, Control of magnetic states and spin interactions in bilayer CrCl3 with strain and electric fields: an ab initio study, Sci. Rep.**13**, 5336 (2023).
[32] W. Zhang, B. Jiang, L. Wang, Y. Fan, Y. Zhang, S.Y. Yu, G.B. Han, G.L. Liu, C. Feng, G.H. Yu, S.S. Yan, and S. Kang, Enhancement of Interfacial Dzyaloshinskii-Moriya Interaction: A Comprehensive Investigation of Magnetic Dynamics, Phys. Rev. Appl.**12**, 064031 (2019).
[33] M. V. Sapozhnikov, R. V. Gorev, E. V. Skorokhodov, N. S. Gusev, A. V. Sadovnikov, and O. G. Udalov, Zigzag domains caused by strain-induced anisotropy of the Dzyaloshinskii-Moriya interaction, Phys. Rev. B**105**, 024405 (2022).
[34] M. Buzzi, R. V. Chopdekar, J. L. Hockel, A. Bur, T. Wu, N. Pilet, P. Warnicke, G. P. Carman, L. J. Heyderman, and F. Nolting, Single Domain Spin Manipulation by Electric Fields in Strain Coupled Artificial Multiferroic Nanostructures, Phys. Rev. Lett. **111**, 027204 (2013).
[35] G.Yu, Z. Wang, M. Abolfath-Beygi, C. He, X. Li, K. L. Wong, P. Nordeen, H. Wu, G. P. Carman, X. Han, I. A. Alhomoudi, P. K. Amiri, and K. L. Wang, Strain-induced modulation of perpendicular magnetic anisotropy in Ta/CoFeB/MgO structures investigated by ferromagnetic resonance, Appl. Phys. Lett. **106**, 072402 (2015).
[36] X. Li, D. Carka, C. Liang, A. E. Sepulveda, S. M. Keller, P. K. Amiri, G. P. Carman, and C. S. Lynch, Strain-mediated 180° perpendicular magnetization switching of a single domain multiferroic structure, J. Appl. Phys.**118**, 014101 (2015).
[37] J. H. Park, H. M. Jang, H. S. Kim, C. G. Park, and S.G. Lee, Strain-mediated magnetoelectric coupling in BaTiO$_3$-Co nanocomposite thin films, Appl. Phys. Lett. **92**, 062908 (2008).
[38] J.-M. Hu, Z. Li, J. Wang, and C. W. Nan, Electric-field control of strain-mediated magnetoelectric random access memory, J. Appl. Phys.**107**, 093912 (2010).

[39] I.O. Gorshkov, R.V. Gorev, M.V. Sapozhnikov, O.G. Udalov, "DMI-gradient-driven skyrmion motion", ACS Applied Electronic Materials **4**, 3205 (2023)

[40] D.A. Tatarskiy, N. S. Gusev, V. Yu. Mikhailovskii, Yu. V. Petrov, and S. A. Gusev, Control over the Magnetic Properties of Co/Pt-based Multilayered Periodical Structures, Technical Physics, **64**, 1584 (2019)

[41] K. Di, V. L. Zhang, H. S. Lim, S. C. Ng, M. H. Kuok, J. Yu, J. Yoon, X. Qiu, and H. Yang, Direct Observation of the Dzyaloshinskii-Moriya Interaction in a Pt/Co/Ni Film, Phys. Rev. Lett. **114**, 047201 (2015).

[42] A.S. Borovik-Romanov, N.M. Kreines, Brillouin-Mandelstam scattering from thermal and excited magnons, Physics Reports **81**, 351 (1982)

[43] J.-H. Moon, S.-M. Seo, K.-J. Lee, K.-W. Kim, J. Ryu, H.-W. Lee, R. D. McMichael, and M. D. Stiles, Phys. Rev. B **88**, 184404 (2013).

[44] L. Sun, R. X. Cao, B. F. Miao, Z. Feng, B. You, D. Wu, W. Zhang, A. Hu, and H. F. Ding, Phys. Rev. Lett. **110**, 167201 (2013).

[45] C. Eyrich, W. Huttema, M. Arora, E. Montoya, F. Rashidi, C. Burrowes, B. Kardasz, E. Girt, B. Heinrich, O. N. Mryasov, M. From, and O. Karis, J. Appl. Phys. **111**, 07C919 (2012).

[46] T. Wu, P. Zhao, M. Bao, A. Bur, J. L. Hockel, K. Wong, K. P. Mohanchandra, C. S. Lynch, and G. P. Carman, Domain engineered switchable strain states in ferroelectric (011) [Pb(Mg$_{1/3}$Nb$_{2/3}$)O$_3$]$_{(1-x)}$-[PbTiO$_3$]$_x$ (PMN-PT, x≈0.32) single crystals, J. Appl. Phys. **109**, 124101 (2011).

[47] O.G. Udalov and I.S. Beloborodov, Strain-dependent Dzyaloshinskii-Moriya interaction in a ferromagnet/heavy-metal bilayer, Physical Review B **102** (13), 134422 (2020).

[48] O.G. Udalov and M.V. Sapozhnikov, Orientation and internal structure of domain walls in ferromagnetic films with anisotropic Dzyaloshinskii-Moriya interaction, J. Magn. Magn. Mater. **519**, 167464 (2021).

[49] K.K. Rajan, M. Shanthi, W.S. Chang, J. Jin, L.C. Lima, Dielectric and piezoelectric properties of [001] and [011]-poled relaxor ferroelectric PZN–PT and PMN–PT single crystals, Sensors and Actuators A 133 110–116 (2007).

[50] L. Angeloni, D. Passeri, M. Reggente, D. Mantovani, and M. Rossi, Removal of electrostatic artifacts in magnetic force microscopy by controlled magnetization of the tip: Application to superparamagnetic nanoparticles, Sci. Rep. **6**, 26293 (2016).